\documentclass[aps,prl,twocolumn,groupedaddress,floatfix]{revtex4}
\usepackage{palatino,color}
\usepackage{graphicx}
\usepackage{amsmath}
\usepackage{natbib}

\begin{document}

\title{Improved Method for Detecting Local Discontinuities in CMB data by Finite Differencing}
\author{Jude Bowyer$^{*}$}
\author{Andrew H. Jaffe$^{\dag}$}
\affiliation{Astrophysics Group, Blackett Laboratory, Imperial College, \\ Prince Consort Road, London, SW7 2BZ, U.K.}

\begin{abstract}
An unexpected distribution of temperatures in the CMB could be a sign of new physics. In particular, the existence of cosmic defects could be indicated by temperature discontinuities via the Kaiser-Stebbins effect. In this paper, we show how performing finite differences on a CMB map, with the noise regularized in harmonic space, may expose such discontinuities, and we report the results of this process on the 7-year Wilkinson Microwave Anisotropy Probe data.

\end{abstract}

\maketitle

  The Kaiser-Stebbins effect[1], a cosmological manifestation of a discontinuous temperature gradient, is a well-known result of the presence of defects in the CMB such as cosmic strings[2]. The magnitude of the gradient relates to the string tension $G\mu$ by
\begin{equation}\frac{\delta T}{T}=8\pi G\mu\alpha_{s}\end{equation}
where $\alpha_{s}$ contains the dynamical and observational information of the string such as the Lorentz factor and the orientation of the string segment; we will assume $\alpha_{s}=1$. Current limits are of order $G\mu\lesssim10^{-6}$[3].
\\  One method of detecting such a discontinuity is by looking for characteristic signals in the derivatives of the CMB temperature field.  Real-space derivative operators produce significant small-scale noise, limiting their utility for cosmology. This is particularly relevant when searching
for highly localized sources such as local defects. The noise properties of harmonic and other transform-based methods, such as wavelets, tend to be well behaved up to the Nyquist limit but these transforms act to smooth out discontinuities. However, it is the different error characteristics of these two types of approaches (with respect to each other) in the presence of discontinuous signals that suggest one might calculate the difference of the derivative
maps created by each method in order to isolate discontinuous signals.
\\ When calculating derivatives over a pixelated sampling grid, a finite difference scheme [4] is both popular and effective. It can be shown that a given finite difference scheme is related to an underlying interpolating polynomial [5]; this being the case, the scheme will perform poorly when in the presence of functions which are not well modeled by such a polynomial, due to Runge’s phenomenon [6]. This is true for discontinuous signals, with the error in the resultant derivative scaling with both the order of the differencing scheme and the severity (the step size) of the discontinuity. These errors can then be used as a crude detector of discontinuities.
\\ In the following text, a method to detect discontinuities and the results of an application to the Wilkinson Microwave Anisotropy Probe (WMAP) 7-year internal linear combination (ILC) CMB temperature anisotropy map are reported. The ILC map is known to suffer from systematic noise at multipoles $l\gtrsim100$, which we further confirm with the detection of a relic of the galaxy boundary. This is performed using the MASQU software package [5], which was initially designed for clean masked CMB polarization mode separation [7] on the popular HEALPIX [8] spherical grid, utilizing real-space differential operators which act on the basis-dependent orthogonal Stokes’ linear polarization parameters $Q$ and $U$ to produce the scalar and pseudoscalar $e$ and $b$ fields, real-space analogues of the $E$ and $B$ modes. In the past, the signal-to-noise ratio in the Stokes’ parameters has been too low for such a method; projects such as the Planck Surveyor [9] will go some way toward rectifying that. In the present case, derivative operations will be performed on the temperature anisotropy map instead. The derivatives of a field with respect to coordinate basis $\hat{n}$ at each point on a pixelated grid can be computed by the finite difference approximation [4]

\begin{equation}\partial_{\hat{n}}F_{i}\approx\sum_{j}w_{ij}^{(\hat{n})}F_{j}\end{equation}

where the $w_{ij}$ are numerical weights attached to a sample of surrounding pixels (the ‘‘pixel stencil’’). The weights are calculated by inverting a matrix $V$ at each pixel, whose elements depend on the positions of the stencil pixels. In the following, $n$th-order calculations \begin{math}O_{n}\end{math} refer to calculations using a square \begin{math}(n+1)^{2}\end{math} pixel stencil. Figure 1 shows how the results vary with both step size and stencil size, for the $E/B$ mode-separation. Note that the error increases both in magnitude and in range for larger stencils in the presence of a discontinuity, whereas it would decrease for a function well-modeled by a polynomial.

\begin{figure}\centerline{\includegraphics[height=50mm]{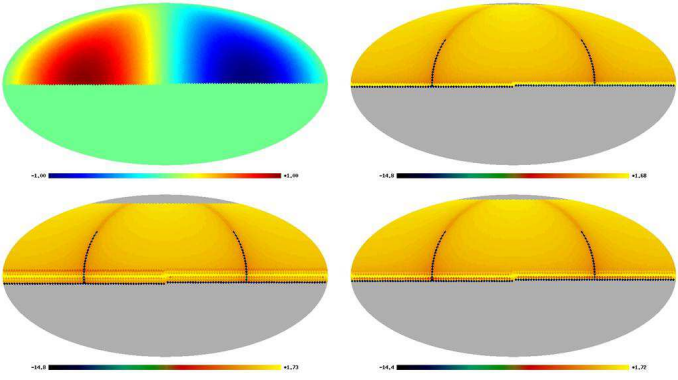}}\caption[]{The effect of step size and stencil size on our derivatives. We calculate the pseudoscalar $\nabla^{4}b$ on a function \begin{math}Q(\theta,\phi)=\sin\theta\cos\phi, U=0\end{math}, with a cut-off to \begin{math}Q(\theta,\phi)=0\end{math} at the equator (top-left diagram), where $b$ is related to the polarization B-mode by $b(\hat{n})=\sum_{lm}\sqrt{\frac{(l-2)!}{(l+2)!}}a^{B}_{lm}Y_{lm}(\hat{n})$. Since there is no $U$ signal, the $b$ signal should be zero everywhere. Clockwise from top-right diagram: $b$ fields for $O_{2},O_{4}$ and $O_{6}$ calculations. The map resolution here is $N_{side}=32$, while the scale is logarithmic.}\end{figure}

	The ILC map[10] is an $N_{side}=512$ resolution temperature aniostropy map, created by the WMAP team by taking weighted combinations of the WMAP band-limited data (bands $K$,$Ka$,$Q$,$V$ and $W$), with special attention paid to the masked region corresponding to the galactic plane. For our analysis of the ILC map, it is useful to calculate the Laplacian of the field, which corresponds to a simple power multiplier
\begin{equation}\nabla^{2}\rightleftharpoons-l(l+1)\end{equation}
when applied to the harmonic space reconstruction of a spherical function
\begin{equation}F(\theta,\phi)=\sum_{l=0}^{\infty}\sum_{m=-l}^{l}a^{f}_{lm}Y_{lm}(\theta,\phi),\end{equation}
calculated explicitly as
\begin{equation}\nabla^{2}F(\theta,\phi)=\sum_{l=0}^{\infty}\sum_{m=-l}^{l}-l(l+1)a^{f}_{lm}Y_{lm}(\theta,\phi)\end{equation}
where the $Y_{lm}$ are the ordinary spherical harmonics.
\\	From here on in, we refer to the method of taking a HEALPIX map, generating the $a_{lm}$s via the iterative HEALPIX $map2alm$ technique[11], multiplying these coefficients by the factor in Eq. (3), and creating a Laplacian map by the HEALPIX $alm2map$ technique as merely ‘‘the spectral method.’’ The spectral method will also expose any conspicuous boundaries, and suffers from Gibbs' phenomenon[12,13], similar to that of Runge. Further, the iterative HEALPIX method used to recreate the $a_{lm}$s is of reasonable but nevertheless limited accuracy. These errors will also propagate through to the Laplacian map.
Since the interpolating polynomial and the spherical harmonics are not identical, a discontinuity should be enhanced by calculating the map of differences between the finite difference Laplacian and the spectral Laplacian. This is then particularly useful when the discontinuity is subtle. As a toy model for the calculation, we show the results for a CMB simulation injected with a diamond-shaped discontinuity (i.e. by linearly adding $\delta T/T=100\mu$K to the map at the discontinuity region, see Fig. 2).

\begin{figure}\centerline{\includegraphics[height=50mm]{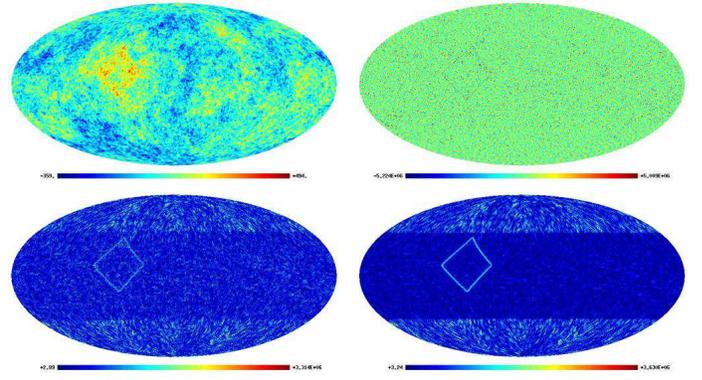}}\caption[]{Top row, left-to-right: CMB simulation ($T_{rms}\approx90\mu$K)+diamond discontinuity, Laplacian map. The boundary of the discontinuity can be spotted by eye. Bottom row, left-to-right: difference maps for $O_2$ and $O_4$ MASQU calculations; the discontinuity boundary is clearly accentuated in these images. The by-eye detection limits for this model in the ILC map are $\delta T/T\sim40\mu$K ($O_{2}$) and $\sim15\mu$K ($O_{4}$), which for strings would correspond to tensions of $G\mu\sim1.5\times10^{-6}$ and $\sim6\times10^{-7}$}\end{figure}

To estimate this calculation, it is noted that the differencing error for derivatives of a function $f(\theta,\phi)$ at pixel $i$ is approximately
\begin{equation}\delta f_{\mathrm{diff},i}(\theta,\phi)\approx\frac{f^{(n+1,p+1)}(\theta,\phi)}{(n+1)!(m+1)!}\mathrm{det}(V)\end{equation}
where $(n,p)$ are the orders of the derivatives in coordinate bases $(\theta,\phi)$, yielding for the full Laplacian
\begin{equation}\begin{split}\delta(\nabla^{2}f)_{\mathrm{diff},i}\approx\left(\csc^{2}\theta \frac{f^{(1,3)}(\theta,\phi)}{3!}+\cot\theta\frac{f^{(2,1)}(\theta,\phi)}{2!}\right.
\\ \hspace{1in}+\left.\frac{f^{(3,1)}(\theta,\phi)}{3!}\right)\mathrm{det}(V).\end{split}\end{equation}

Meanwhile, the function reconstruction error from the Gibbs' phenomena receives contributions from two places; First, the Fourier sum in the limit of $l\rightarrow\infty$ will reconstruct the signal everywhere except at the discontinuity. Secondly, truncation creates overshoot (ringing) in the vicinity of the discontinuity. These errors will be of the form
\begin{equation}\delta f_{\mathrm{recon},i}=\left|f(\theta,\phi)-\sum_{lm}^{\infty}a^{f}_{lm}Y_{lm}(\theta,\phi)\right|\end{equation}
and
\begin{equation}\delta f_{\mathrm{trun},i}=\left|\sum_{lm}(a^{f}_{lm}-a^{g}_{lm}) Y_{lm}(\theta,\phi))\right|\end{equation}
where $g$ is a truncated approximation to the original function $f$, cut off at $l_{max}$
\begin{equation}g(\theta,\phi)=\sum_{lm}^{l_{max}}a^{f}_{lm}Y_{lm}(\theta,\phi)\end{equation}
An analysis of this type on the sphere has been carried out for longitudinal ringing via the Gegenbauer polynomials in [14]. Finally, there is an error contribution $\delta f_{\mathrm{hpix}}$ which accounts for any small remaining errors accumulated from the HEALPIX routines. In calculating the Laplacian of the field, the full harmonic function errors $\delta f_{\mathrm{spec}}$ will have additional scaling as $\sim l^{2}$. Our method thus calculates
\begin{equation}\delta(\nabla^{2}f)_{i}\approx\delta(\nabla^{2}f)_{\mathrm{diff},i}-\nabla^{2}(\delta f_{\mathrm{spec},i}).\end{equation}
at each pixel position $i$. The noise characteristics (Fig. 3) of the absolute difference map accounts for why a discontinuity can be seen by eye in the toy model absolute difference map; the $\sim l^{2}$ scaling in either Laplacian map is approximately canceled, leaving white noise on small scales.

\begin{figure}\centerline{\includegraphics[height=60mm,width=85mm]{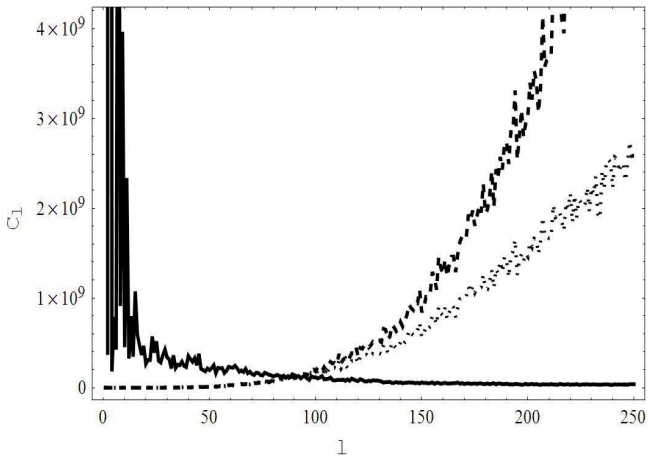}}\caption[]{Thick line: power spectrum for the $O_{2}$ absolute difference map calculated over a map synthesized from a Gaussian random sample of $a_{lm}$s provided by the ($\sigma=1\mu$K) HEALPIX random number generator facility. The dashed and dotted lines are the spectra for the spectrally-calculated and $O_{2}$ finite-differenced Laplacian maps respectively. The discrepancy between the dashed and dotted lines decreases with an increased number of sampling points in the finite-differencing calculations. The power spectrum $C_{l}^{|\nabla^{2}T_{\mathrm{spec}}-\nabla^{2}T_{\mathrm{F.D.}}|}$ of the map differences has a white spectrum on small scales due to the strong correlation of noise between the maps and a boost on large scales due to the HEALPIX geometry in the polar cap (see [5]).}\end{figure}

  The results for the software performance on the ILC data are shown in Fig. 4. Taking the absolute difference between the finite-differenced ILC map and spectrally calculated ILC map reveals the galaxy mask boundary very clearly, regardless of the smoothing used in the ILC map construction[10].

	 \begin{figure}\centerline{\includegraphics[height=50mm]{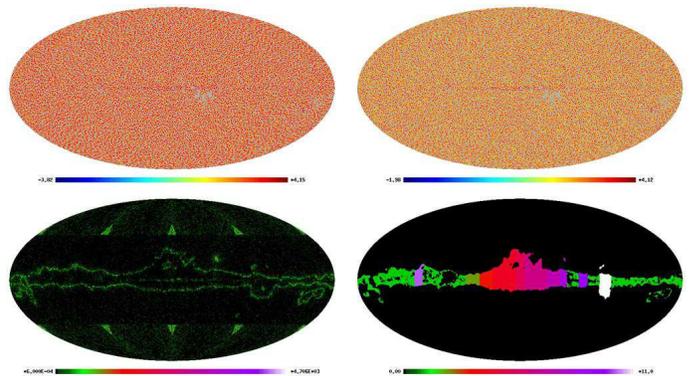}}\caption[]{Top row: ILC log Laplacian maps from the (left diagram) $map2alm$ method and (right diagram) finite differencing at $O_{2}$. Bottom row: (left diagram) the difference map between the above. The galaxy mask is quite apparent, as are a couple of signals which are likely galactic foreground sources. The triangular parts in the polar cap are relics of the calculation technique and HEALPIX geometry. The last plot is the mask map,with ILC technique cuttings shown.}\end{figure}

The ILC map is not directly useful for cosmological parameter estimation; instead, it is used as an approximately noise-free CMB map in order to isolate galactic foregrounds and generate a zeroth-order noise template [15] which is then supplemented by a maximal entropy method. In this sense the impact of removing the boundary discontinuity region for analyses is unlikely to seriously effect the cosmological implications of the WMAP results. Instead of implementing the complicated WMAP-type analysis to reconstruct the angular power spectrum, the boundary is isolated in a separate mask. One can then estimate the contribution of the masked region to the ILC power spectrum by using the relation between the pseudo-$C_l$ spectrum  $\tilde{C}_l$ and the theoretical power spectrum given below:

\begin{equation}\tilde{C}_{l}=\sum_{ll'}K_{ll'}C_{l'}\end{equation}

via a mode-coupling matrix $K_{ll'}$. More typically, this technique is used in MASTER-type calculations[16] (a Monte Carlo-type power spectrum estimator, such as SPICE[17]) since the equation above properly recovers only the Monte Carlo average of the spectral signal. This is performed for a range of boundary thicknesses (2-,4- and 6-pixel mask thicknesses), since the smoothing kernel corresponds to 1.5$\deg$[18] which is not rigorously followed in the code used to create the boundary mask.

	 \begin{figure}\centerline{\includegraphics[height=50mm]{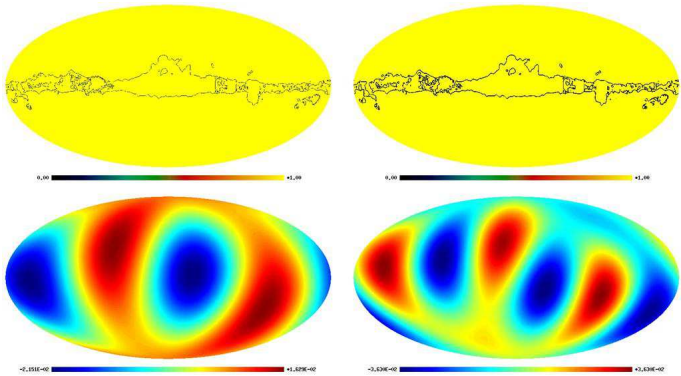}}\caption[]{Top row: Boundary masks used, of varying thickness. Bottom row: alteration in the quadrupole (left diagram) and octupole (right diagram) power from the smaller boundary mask. The alignment of the quadrupole and octupole is essentially unchanged.}\end{figure}

	It can be seen in Fig. 6 that the impact of the boundary region is small; however, as cosmological experiments probe the microwave sky with higher precision and theory discrimination becomes more sensitive to subtle differences in predicted power spectra, it may be necessary to ensure that excess power from the mask boundary does not bias results. Secondly, given recent claims of detections of non-Gaussianity[19,20] it would certainly be necessary to account for the boundary region in any calculation of the nonlinear term $f_{NL}$[21]. The effect of the boundary region on a number of the so-called anomalies[22] has also been calculated, with no significant divergence from their manifestation when the boundary region is included (Fig. 5).
	
	 \begin{figure}\centerline{\includegraphics[height=60mm]{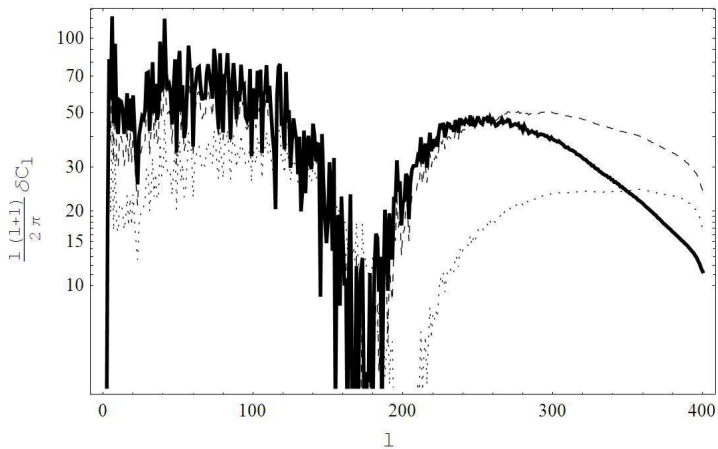}}\caption[]{Differences between the boundary masked and unmasked ILC power spectra --- the lines correspond to the thickest mask (thick line), thinnest mask (dotted line) and intermediate mask (dashed line). The dip in power at $l\sim180$ corresponds roughly to the Gaussian smoothing scale across the galaxy boundary in the construction of the ILC map, so the difference between ILC and mode-uncoupled masked spectra at the mask scale is small.}\end{figure}	
	
	One can continue the search for anomalies by excluding the galaxy region precalculation and generating a new Laplacian-difference map. For firm statistical analysis (eschewing any by-eye method), the needlet approach[23] is particularly suitable to decomposing a discontinuity into power due to its localization properties, while the Canny algorithm[24] is suited to edge detection. Further, it should be noted that there may be correlations at the discontinuity boundary with the Stokes' $Q$ and $U$ parameters, since cosmic strings have a vector $B$-mode signal[25] in real space analogous to the Kaiser-Stebbins effect in the temperature case. These issues will be discussed in a future paper.
\\In conclusion, the method proposed finds no immediate evidence for anomalies such as strings, with the caveat that only lower-order calculations have been performed, coupled with by-eye detection. This finding may change with higher-order calculations and a more rigorous examination using the needlet algorithm. Also, since the signal in a pixel is an average over signals within the pixel area, it is possible that the results of the Planck Surveyor with an improved resolution of \begin{math}N_{side}=2048\end{math}, or more local balloon surveys such as the E and B Experiment [26] and their successors, will test these results. The method did encounter a success: the detection of the ‘‘smoothed’’ galaxy boundary in the map. It was found that the effect of excluding the boundary from the map was negligible, but that its mere detection might warrant added care in future surveys when performing map synthesis.

\begin{acknowledgments}
Thanks go to H. V. Peiris for promoting the needlet approach. This work was supported by the STFC.
\end{acknowledgments}

\end{document}